\documentclass[useAMS,usenatbib,usegraphicx,rotating,color,multirow,upgreek]{aa}  
\usepackage{graphicx}
\usepackage{txfonts}
\usepackage{epsfig}
\usepackage{amsmath} 
\usepackage{rotating}           
\usepackage{color}     
\usepackage{graphicx}
\usepackage{upgreek} 

\def\Lo{L$_\odot$}
\def\Mo{M$_\odot$}
\def\Mom{{\rm M}_\odot} 
\def\Moy{M$_\odot$~yr$^{-1}$}

\def\ccm {$\hbox{{\rm cm}}^{-3}$}    
\def\scm  {$\hbox{{\rm cm}}^{-2}$}    
\def \AL {$\alpha $}     
\def \HI {H{\sc \,i}}
\def \WpHz {W Hz$^{-1}$}

\def\lapp{\ifmmode\stackrel{<}{_{\sim}}\else$\stackrel{<}{_{\sim}}$\fi}
\def\gapp{\ifmmode\stackrel{>}{_{\sim}}\else$\stackrel{>}{_{\sim}}$\fi}

\begin{document} 
\title{Ultra-violet photo-ionisation in far-infrared selected sources}
 \author{S. J. Curran\inst{1}
           \and
           S. W. Duchesne\inst{2}
         }
  \institute{School of Chemical and Physical Sciences, Victoria University of Wellington, PO Box 600, Wellington 6140, New Zealand\\
  \email{Stephen.Curran@vuw.ac.nz} 
        \and
        International Centre for Radio Astronomy Research (ICRAR), Curtin University, Bentley, WA 6102, Australia}

\abstract{
It has been reported that there is a deficit of stellar heated dust, as evident from the lack of far-infrared (FIR)
  emission, in sources within the {\em Herschel}-SPIRE sample with X-ray luminosities exceeding a ``critical value'' of
  $L_{\mathrm{X}} \sim10^{37}$~W. Such a scenario would be consistent with the suppression of star formation by the AGN,
  required by current theoretical models.  Since absorption of the 21-cm transition of neutral hydrogen (\HI), which
  traces the star-forming reservoir, also exhibits a critical value in the ultra-violet band (above ionising photon
  rates of $Q\approx3\times10^{56}$~s$^{-1}$), we test the SPIRE sample for the incidence of the detection of
  $250$~$\mu$m emission with $Q$.  The highest value at which FIR emission is detected above the SPIRE confusion limit
  is $Q=8.9\times10^{57}$~s$^{-1}$, which is $\approx30$ times that for the \HI, with no critical value apparent. Since
  complete ionisation of the neutral atomic gas is expected at $Q\gapp3\times10^{56}$~s$^{-1}$, this may suggest that
  much of the FIR must arise from heating of the dust by the AGN.  However, integrating the ionising photon rate of each
  star over the initial mass function, we cannot rule out that the high observed ionising photon rates are due to a
  population of hot, massive stars.
}
  

   \maketitle
%

\section{Introduction}
\label{intro}

Feedback between active galactic nuclei (AGN) and their host galaxies is complex, with the central engine ionising the very
material which feeds it (e.g. \citealt{dsh05,fab12}).  The ionisation can arise from powerful outflows into the
surrounding neutral medium (jet/radio-mode, e.g. \citealt{cmm+12,ful+12}), as well as by ultra-violet (\citealt{sr98})
and X-ray (\citealt{fab99a}) radiation emanating from the accretion disk surrounding the super-massive black hole
(radiative/quasar-mode, \citealt{hec07,hb14}). Both mechanisms will suppress star formation through ionisation and
heating of the neutral gas, this suppression being required by current theoretical models to reproduce the observed
properties of active galaxies (e.g. \citealt{csw+06}).  Observational evidence of this has recently been claimed, where
powerful AGN, as evident through their X-ray luminosity, lack $\lambda=250$~$\mu$m far-infrared (FIR) emission over the
redshift range $1 < z < 3$. In the rest-frame this corresponds to wavelengths of $63 - 125~\mu$m, which trace the dust heated
by stars (e.g. \citealt{dh02}), and so the absence of FIR emission is interpreted as the suppression of star formation in the
 X-ray luminous sources \citep{psv+12}.

Powerful AGN are also extremely bright in the UV, which is redshifted into the optical band at $z\gapp3$, allowing
quasi-stellar objects (QSOs) to be visible to ground-based telescopes over much of the observable
Universe. Observational evidence for high UV luminosities rendering the star-forming material undetectable in the hosts
of radio-loud QSOs (quasars) was suggested by the exclusive non-detection of the 21-cm transition of neutral hydrogen
(\HI) above a ``critical'' UV luminosity of $L_{\rm UV}\sim10^{23}$~\WpHz \citep{cww+08}.  \HI\ 21-cm absorption traces
the cool, star-forming, component of the gas and a subsequent model of a quasar placed within an exponential gas disk
found that this luminosity, which corresponds to $Q\sim3\times10^{56}$ ionising ($\lambda \leq912$ \AA) photons s$^{-1}$, is
just sufficient to ionise all of the neutral gas in the Milky Way \citep{cw12}. Given that this is a large spiral galaxy, complete
ionisation of the neutral gas would explain why \HI\ has never yet been detected in sources with these UV luminosities
\citep{cww+08,cwm+10,cwsb12,cwt+12,caw+16,cwa+17,chj+17,chj+19,ace+12,gd11,gmmo14,akk16,akp+17,ak17,cd18,gdb+15}. 
Since  \HI\ 21-cm absorption traces the reservoir for star formation, we may also expect the 
250~$\mu$m emission to be absent in the sources above the critical UV ionising photon rate, a possibility we investigate here.

\section{Analysis}

\subsection{Source matching}

\begin{figure}
\centering 
\includegraphics[angle=270,scale=0.5]{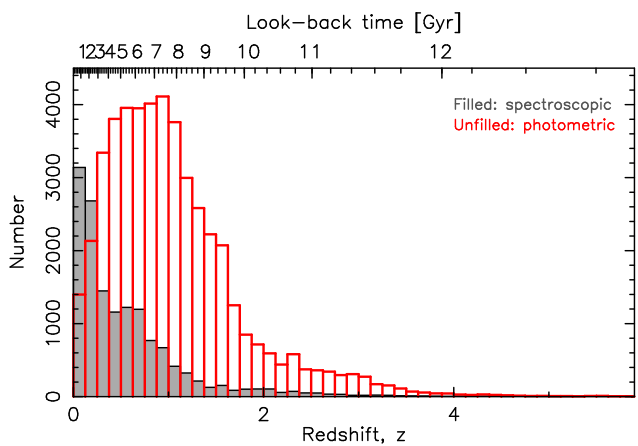}
\caption{The distribution of the spectroscopic (filled histogram) and the photometric redshifts (unfilled) for the
62\,073 matched SPIRE sources.}
\label{z-histo}
\end{figure}

The {\em Herschel Multi-tiered Extragalactic Survey} (HerMES, \citealt{oba+12}), utilised by \citet{psv+12}, covers a
$\sim 340$~deg$^2$ field, operating at wavelengths of $\lambda=250$, 350 and 500~$\mu$m ({\em Spectral and Photometric
  Imaging Receiver}, SPIRE) and 100 and 160~$\mu$m ({\em Photodetector Array Camera and Spectrometer}, PACS).  We
constructed a subset of extragalactic sources within the SPIRE dataset using the \textsc{StarFinder} source-finder software
\citep{dbb+00}, with cross-identification being performed between the single-band catalogues \citep{rok+10}, resulting
in a merged catalogue of 340\,968 sources.  We then searched the {\em NASA/IPAC Extragalactic Database} (NED) for
sources within 6~arcsec of the SPIRE positions (as per \citeauthor{psv+12}, see Appendix~\ref{sec:x-ray}), which were
cross-matched according to the minimum separation between sources, while ensuring no duplicate matches.  After filtering
out SPIRE sources without a counterpart nor a redshift, there were 62\,073 sources remaining. Of these, 14\,457 have
spectroscopic redshifts (Fig.~\ref{z-histo}).

\subsection{Photometry fitting}
\label{sec:pf}

\subsubsection{Ultra-violet}

In order to obtain the UV luminosities, we queried NED, the {\em Wide-Field Infrared Survey Explorer} (WISE,
\citealt{wem+10}), the {\em Two Micron All Sky Survey} (2MASS, \citealt{scs+06}) and the {\em Galaxy Evolution Explorer}
(GALEX data release GR6/7)\footnote{http://galex.stsci.edu/GR6/\#mission} databases.  Each flux density measurement,
$S_{\nu} $, was corrected for Galactic extinction \citep{sfd98}, before being converted to a specific luminosity at the
source-frame frequency, via $L_{\nu}=4\pi \, D_{\rm L}^2\,S_{\nu}/(z+1)$, where $D_{\rm L}$ is the luminosity distance
to the source.

The ionising photon rate is defined as $Q\equiv \int^{\infty}_{\nu_0}(L_{\nu}/h\nu)d{\nu}$ \citep{ost89}, where $\nu_0 =
3.29\times10^{15}$~Hz for the ionisation of neutral hydrogen. For the fitting, we require at least three
$\nu\gapp10^{15}$~Hz photometry points to which we fit a power-law, giving $\log_{10}L_{\nu} =
\alpha\log_{10}\nu+ {\cal C} \Rightarrow L_{\nu} = 10^{\cal C}\nu^{\alpha}$, where \AL\ is the spectral index and ${\cal
  C}$ the intercept. This gives the ionising photon rate as
\[
Q = \frac{10^{\cal C}}{h}\int^{\infty}_{\nu_0}\nu^{\alpha-1}\,d{\nu} = \frac{10^{\cal C}}{\alpha h}\Bigg[\nu^{\alpha}\Bigg]^{\infty}_{\nu_0} = \frac{-10^{\cal C}}{\alpha h}\nu_0^{\alpha},
\]
where $\alpha < 0$.

\subsubsection{Far-infrared}
\label{FIR}

\begin{figure}
\centering 
\includegraphics[angle=270,scale=0.52]{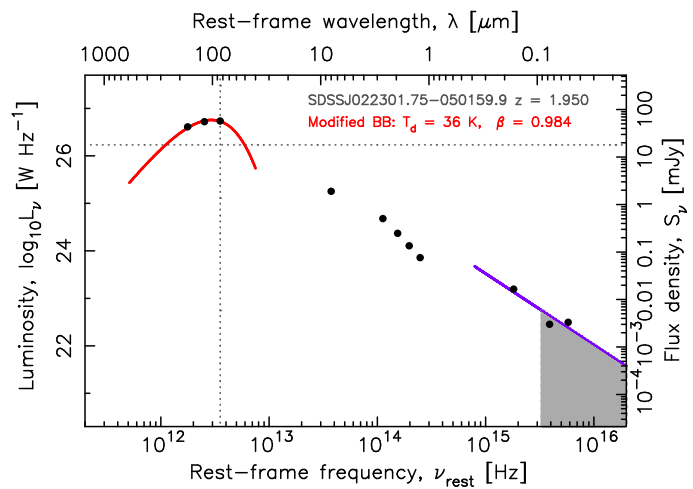} 
\vspace*{5mm}
\caption{Example of the FIR and UV photometry fits. The vertical line shows the observed-frame 250~$\mu$m frequency
and the horizontal line the SPIRE confusion limit of $3\sigma_{\rm conf}=18$ mJy. The shaded region shows the 
$\lambda \leq 912$~\AA\ band over which the ionising photon rate is derived.}
\label{phot-ex}
\end{figure}

Since we were also interested in the effect of the UV continuum on the dust temperature, 
we added the photometric data of the three SPIRE bands to that above, where we only included the data detected at
$S_{\nu}> 3\sigma_{\rm conf}$, where $\sigma_{\rm conf}=6$~mJy is the SPIRE confusion limit \citep{nsl+10,swo+12}. For
$T_{\rm dust}\lapp h\nu/k$ \citep{yof+09}, where $h$ is the Planck constant and $k$ the Boltzmann constant, we fitted a
modified blackbody (Fig. \ref{phot-ex}) spectrum of the form
\[
S_{\nu} \propto \frac{\nu^{3+\beta}}{e^{h\nu/kT_{\rm dust}}-1},
\]
where $\beta$ is the spectral emissivity index (e.g. \citealt{cas12}). 

\begin{figure}
\centering 
\includegraphics[angle=270,scale=0.44]{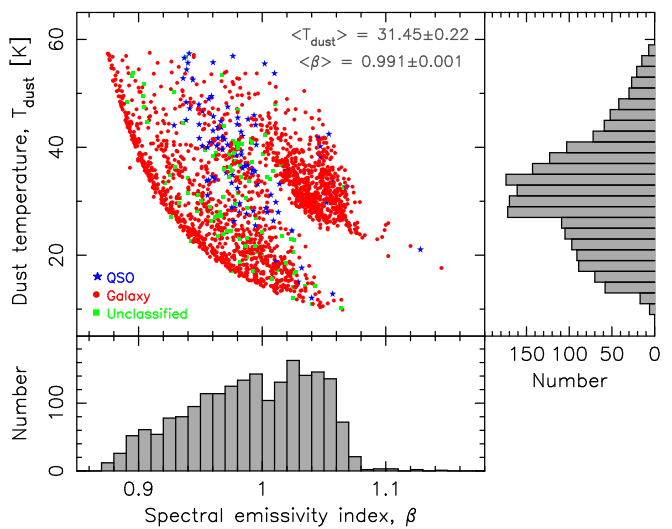}
\caption{The distribution of spectral emissivity index and dust temperature, where $T_{\rm
    dust}\leq1.2\times10^{12}(z+1)h/k$, giving a sample of $n=2013$.  The shapes represent the NED classification --
  stars for QSOs, circles for galaxies and squares for unclassified.}
\label{beta-T}
\end{figure}

For this, we used a non-linear least-squares fitting of  both $\beta$ and $T_{\rm dust}$
simultaneously, giving the distribution shown in Fig.~\ref{beta-T}.
This yields a mean dust temperature of $\left<T_{\rm dust}\right> = 31$~K and a mean spectral emissivity index of
$\left<\beta\right>=0.99$. This is considerably lower than the canonical $\beta=1.5$ (\citealt{bbc03,yof+09} and
references therein), although this was used as the initial estimate. Forcing $\beta=1.5$ could not satisfactorily fit
the data, although we do note that $\beta\approx1.0$ may not be unexpected 
\citep{hil83}.

\section{Effects of the ultra-violet continuum}
\subsection{Critical ionising photon rate}
\label{cipr}

As stated in the introduction, our main motivation was to test whether the critical ionising photon rate found
for \HI\ 21-cm absorption also applied to FIR emission, where an apparent critical X-ray luminosity may 
suggest the suppression of star formation. The highest ionising photon rate at which 21-cm absorption
has been detected is $Q=2.5\times10^{56}$~s$^{-1}$, above which there are 87 non-detections, which 
is significant at $5.66\sigma$ (\citealt{chj+19}). 
Since 21-cm absorption traces the cool, neutral gas that fuels the star formation, we may therefore expect
a similar critical rate if the FIR emission is dominated primarily by stellar activity.


Of the sample, there is sufficient UV photometry for 3315 sources, of which 1347 are considered 250~$\mu$m detections
(where $S_{250 \mu{\rm m}}> 18$ mJy, Sect. \ref{FIR}), giving a detection rate of 40.6\%. 
\begin{figure*}
\centering \includegraphics[angle=-90,scale=0.67]{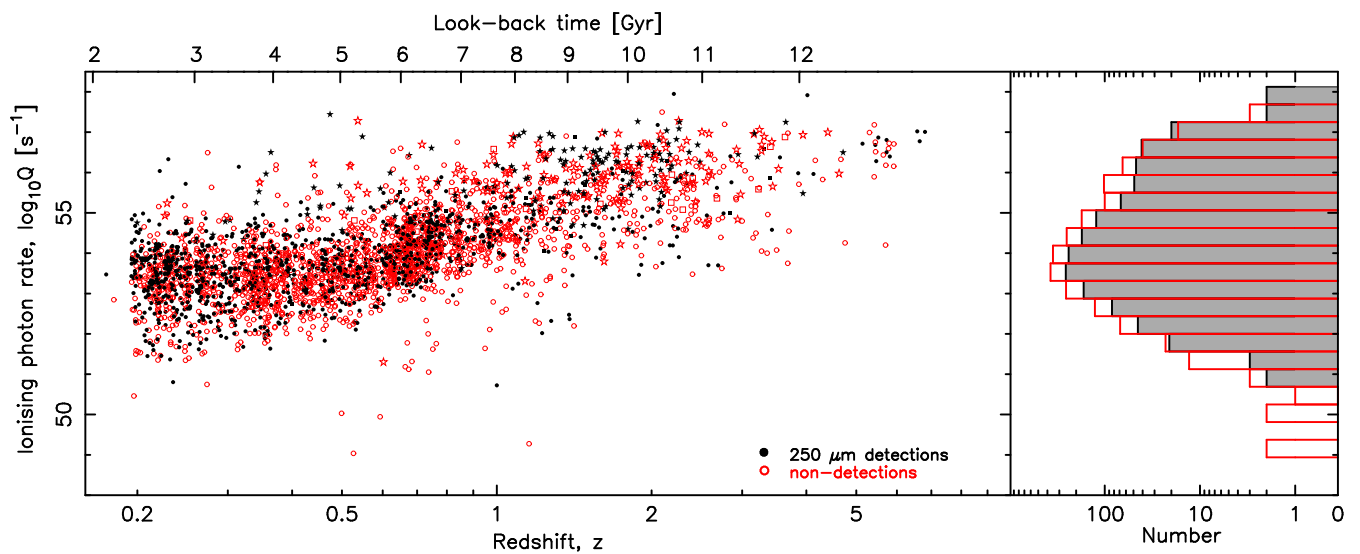}
\caption{The ionising ($\lambda \leq912$ \AA) photon rate versus redshift for the sources with spectroscopic redshifts. 
The filled symbols/histogram represent the FIR detections and the unfilled the non-detections. 
As per Fig. \ref{beta-T}, the shapes represent the NED classification -- stars for QSOs, circles for galaxies and 
squares for unclassified.}
\label{Q-z}
\end{figure*}
As seen from Fig.~\ref{Q-z}, unlike for 21-cm absorption, the detections and non-detections
share the same range of ionising photon rates, with 250~$\mu$m emission being detected up to $Q=8.9\times10^{57}$~s$^{-1}$. This is considerably higher than that observed for \HI\ 21-cm absorption and the model prediction that all of
the neutral gas in a large galaxy is ionised at $Q\gapp3\times10^{56}$ s$^{-1}$ \citep{cw12}. Thus, unlike the X-ray
emission (\citealt{psv+12}, but see Appendix~\ref{sec:x-ray}), we see no distinction between the distribution of FIR
detections and non-detections at high ionising photon rates.

\subsection{Dust heating}
\label{dusttemp}

\begin{figure}
\centering \includegraphics[angle=-90,scale=0.47]{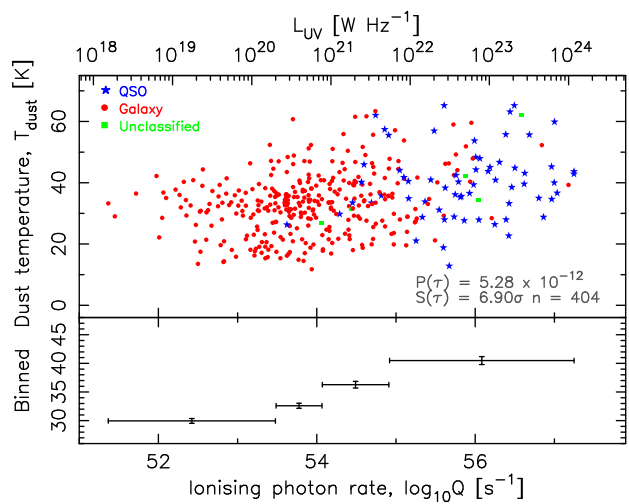}
\caption{The dust temperature versus the ionising photon rate. In the bottom panel the data are shown in equally sized
  bins, where the horizontal bars show the range of points in the bin and the vertical error bars the $1\sigma$
  uncertainty in the mean value.}
\label{temp-Q}
\end{figure}

Interstellar dust has an absorption cross-section which peaks at ultra-violet wavelengths, thus making the re-emitted
far-infrared radiation a sensitive tracer of dust heating.  From the modified blackbody fits to the FIR photometry, we
can also investigate heating of the dust by the ultra-violet emission.  Of the 250~$\mu$m detections, there are 404
sources to which we could fit a modified blackbody and which have sufficient UV photometry
(e.g. Fig.~\ref{phot-ex}). These exhibit a strong correlation between the dust temperature and the ionising photon rate
(Fig. \ref{temp-Q}),
with a generalised non-parametric Kendall-tau test giving a probability of $P(\tau) = 5.28\times10^{-12}$ for the
$T_{\rm dust}-Q$ correlation arising by chance, which is significant at $S(\tau)=6.90\sigma$ assuming Gaussian
statistics.  However, due to the flux limitation introduced by the SPIRE confusion limit, there will be a bias towards
the most FIR luminous, and thus most UV luminous, sources with increasing redshift. Given that for a (modified)
blackbody, the temperature and intensity are not independent (Sect.~\ref{FIR}), this will also lead to an apparent
increase in the dust temperature.

\begin{figure}
\centering \includegraphics[angle=-90,scale=0.47]{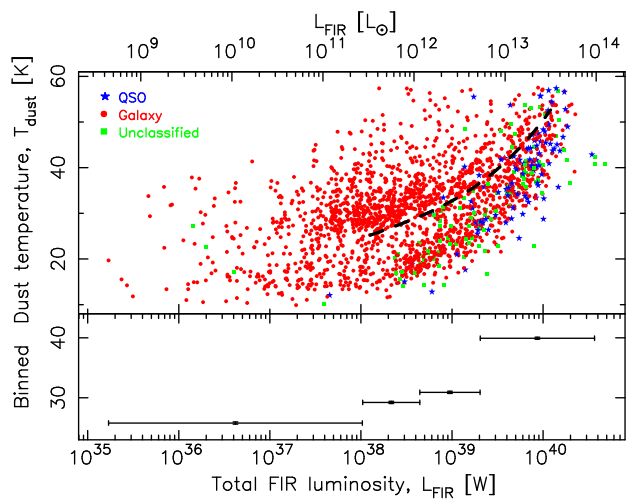}
\caption{The dust temperature versus the total FIR  luminosity. The broken curve shows the fit for the SPIRE-mm sources \citep{rig+11}.}
\label{temp-bolo}
\end{figure}

\begin{figure}
\centering \includegraphics[angle=-90,scale=0.47]{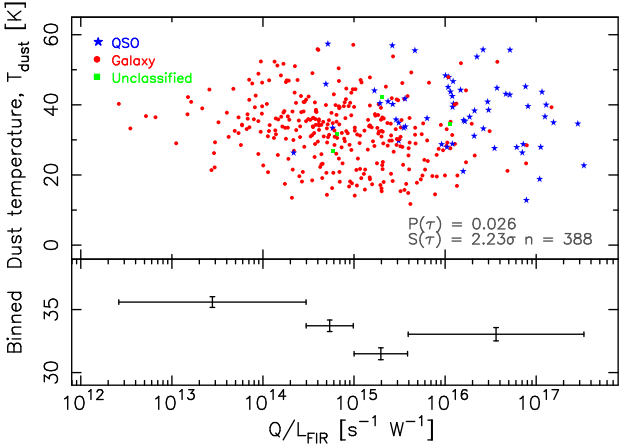}
\caption{The dust temperature versus the ionising photon rate normalised by the total FIR  luminosity.}
\label{temp-ratio}
\end{figure}

In order to correct for this, we obtain the integrated intensity over  40--1000~$\mu$m (\citealt{ygdb07}, cf. Fig.\ref{phot-ex})
from
\[
I_{\rm FIR} = \int_{3.0\times10^{11}\,\text{Hz}}^{7.5\times10^{12}\,\text{Hz}}\frac{2\pi h\nu^{3+\beta}}{c^2} \frac{1}{e^{h\nu/kT_{\rm dust}}-1} d\nu,
\]
from which a specific intensity ($I_{250\,\mu{\rm m}}$) is compared to the corresponding specific luminosity
($L_{250\,\mu{\rm m}}$, e.g. Fig. \ref{phot-ex}) in order to provide the scaling to $L_{\rm FIR}$
(Fig. \ref{temp-bolo}).
From this, we see that many of the luminosities are in excess of $L_{\rm FIR}\gapp10^{12}$~\Lo, 
qualifying the sources as {\em  Ultra-Luminous Infrared Galaxies} (ULIRGs). 
Normalising the ionising photon rate by the FIR luminosity (Fig. \ref{temp-ratio}),
the correlation disappears. This and Fig.~\ref{temp-bolo} therefore indicate that the increase in dust temperature with
redshift\footnote{For example, $T_{\rm dust} = 19-25$~K in near-by galaxies \citep{gka+12}, $T_{\rm dust} =31-36$~K in
  intermediate redshift starburst galaxies and AGN (\citealt{kob+10,mrh+13}) and $T_{\rm dust} \approx40$~K in
  intermediate and high redshift ULIRGs (\citealt{ygdb07} and \citealt{yof+09},
  respectively).}, is due to the flux limitation introducing a bias towards the most FIR (and UV) luminous sources.

\subsection{Stellar versus AGN activity}

The FIR emission is believed to be dominated by stellar heating (\citealt{rl79b} and references therein), whereas the
mid-infrared (MIR, $\lambda=24$~$\mu$m) emission can arise in photon dominated/H{\sc ii} regions (e.g. \citealt{ggj18}),
as well as from AGN heated dust in the circumnuclear torus \citep{hos+10}.\footnote{Invoked by unified schemes of AGN
(e.g. \citealt{ant93,up95}).}  However, the absence of the same critical ionising photon rate, which completely ionises
the neutral atomic gas, and thus suppresses the star formation, at $Q\gapp3\times10^{56}$ s$^{-1}$ \citep{cw12}, may
suggest that much of the FIR emission also arises from dust heated by the AGN (e.g. \citealt{dmld85,cjhb99,nrw+10}).

For a blackbody with $T=5770$~K and a radius of $R_{\odot}=696\times10^8$~m, the ionising photon rate of the Sun is
$Q_{\odot} = 7.9\times10^{35}$~s$^{-1}$. Thus, UV luminosities corresponding to $Q\gapp10^{57}$ s$^{-1}$
(Fig.~\ref{Q-z}) would require $\gapp10^{21}$ solar mass stars, and so, while an older (lower mass)
 population of stars could contribute significantly to the FIR luminosity \citep{cwh+10,gks+12,lcc+13}, these cannot account for the UV luminosity.
In order to explore the stellar masses required, the ionising photon rate of a star is obtained from
the integrated intensity, $I_{\nu}$, 
via
\[
\begin{split}
Q_{\star} h = L_{\rm UV} = 4\pi R_{\star}^2\int^{\infty}_{\nu_0}I_{\nu}d{\nu},\\
\text{ where } I_{\nu}= \frac{2\pi h\nu^3}{c^2} \frac{1}{e^{h\nu/kT_{\star}}-1},
\end{split}
\]
$\nu_0 = 3.29\times10^{15}$~Hz corresponds to $\lambda=912$ \AA, where the hydrogen is ionised, and $T_{\star}$ is the surface temperature of the star.  To obtain the luminosity we
require a value for the surface area of the star ($4\pi R_{\star}^2$), which we estimate from the comparison of the bolometric intensity, 
$I= \int^{\infty}_{0}I_{\nu}d{\nu}$, with the bolometric luminosity obtained from the main sequence using $\log_{10}L_{\star} =
6.50 \log_{10}T_{\star} -24.37$. Examples of the derived properties are given in Table~\ref{stars}.
\begin{table} 
  \caption{The estimated properties of stars of various  temperatures. $L_{\star}$ is the stellar luminosity
    estimated from the main sequence, followed by the radius required to equate this to the bolometric intensity obtained from the Planck function.  The stellar  mass is estimated from the mass--luminosity relation, $L_{\star}/L_{\odot} = (M_{\star}/M_{\odot})^{3.5}$, and the main sequence lifetime, $t_{\rm MS}$, from $t_{\rm MS}\propto 1/M^{2.5}$.
    $N_{\rm stars}$ is the number of stars of this mass according to the Salpeter initial mass function, normalised to a total stellar mass of $3\times10^{12}$~\Mo, followed by the star formation required to maintain this (see main text).}
\begin{tabular}{l | c  c c} 
\hline\hline
\smallskip
$T_{\star}$ [K] & 20\,000 & 35\,000 & 50\,000 \\
\hline
$L_{\rm total}$ [L$_{\odot}$] & 4200  &  $2\times10^5$& $2\times10^6$ \\
Radius [R$_{\odot}$] & 5.4   & 11 & 17\\
Mass [M$_{\odot}$] &  11  &  31 & 60\\
$t_{\rm MS}$ [yr] &  $3\times10^{7}$ & $2\times10^{6}$ & $4\times10^{5}$\\
$Q_{\star}$ [s$^{-1}$]  & $2.7\times10^{46}$& $1.9\times10^{48}$& $3.4\times10^{49}$\\
$N_{\rm stars}$   &  $7.9\times10^{8}$  & $7.0\times10^{7}$  & $1.5\times10^{7}$\\
SFR [\Moy] & $300$   & $1100$ & $2500$\\
\hline
\end{tabular}
\label{stars}  
\end{table} 

\begin{figure}
\centering \includegraphics[angle=-90,scale=0.52]{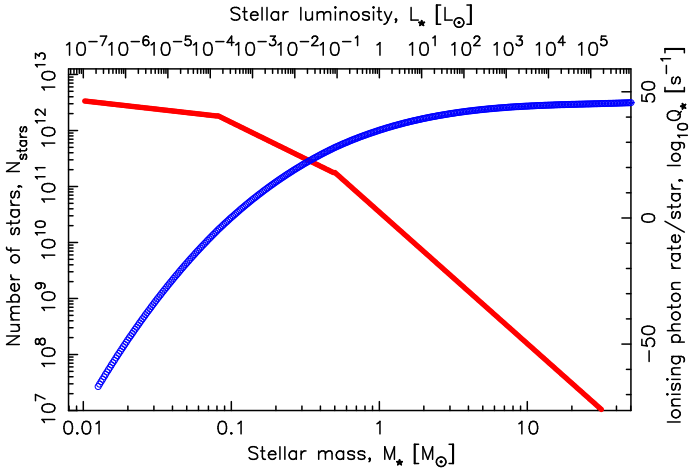}
\caption{The frequency of various stellar masses obtained from the initial mass function, $\xi = \xi_0M^{\alpha}$, 
normalised to give a total stellar mass of $M_{\rm total} = 4\times10^{11}$ \Mo. $\alpha=-0.3$ for $0.01 < \Mom\ < 0.08$, $\alpha=-1.3$ for $0.08 < \Mom\ < 0.5$ and  $\alpha=-2.35$ for $\Mom\ > 0.5$ \citep{sal55,kro01},
The curve (right-hand scale) shows the approximate ionising photon rate per star.}
\label{IMF}
\end{figure}

In order to obtain the total ionising luminosity,  we integrate both the initial mass
function (IMF or $\xi$) and $Q_{\star}$ over the whole mass range, which we 
normalise to the approximate stellar mass of the Milky Way (Fig.~\ref{IMF}).
The total stellar mass is given by $M_{\rm total} = \int_{M_1}^{M_2}M\xi dM$ and is dominated by the low mass stars,
whereas the total ionising photon rate is dominated by the massive stars (cf. Table \ref{stars}). If we truncate the IMF
at $M_{\star} = 25$~\Mo\ ($T_{\star}\approx30\,000$~K, $t_{\rm MS}\approx3\times10^6$~yr) for the maximum possible mass,
we obtain $Q_{\rm total} = 3\times10^{56}$~s$^{-1}$, whereas setting this to $M_{\star} = 100$~\Mo\
($T_{\star}\approx65\,000$~K, $t_{\rm MS}\approx1\times10^5$~yr), gives $Q_{\rm total} = 7\times10^{57}$~s$^{-1}$.

Thus, the total ionising photon rate from the stellar population within the source is very much dependent upon the upper
end of the stellar mass distribution.  Assuming no significant AGN contribution, the highest FIR luminosities of the
sample indicate star formation rates of $\sim10^4$ \Moy\ (e.g. \citealt{ken98}), which we can use to constrain the
upper mass end. In this instance, integrating over all star formation rates\footnote{The specific star
formation rate at a given mass is estimated from SFR\,$=N_{\rm stars} M/t_{\rm MS}$. For example, from Table~\ref{stars},
we expect  $8\times10^8$  stars of $\approx10$~\Mo, which have a lifetime of $t_{\rm MS} \approx 3\times10^{7}$~yr.
This therefore requires  SFR$\,\approx 8\times10^8 \times 10/3\times10^7\approx 300$~\Moy\  to maintain the 
observed luminosity.}, a total of SFR$_{\rm total}\,=1\times10^4$~\Moy\ is reached for a
maximum stellar mass of $M_{\star}\approx40$~\Mo, which gives $Q_{\rm total} =9\times10^{56}$~s$^{-1}$.
However, even at relatively large look-back times, e.g. $\sim10$~Gyr ($z\sim2$), this implies a total stellar mass of 
$\sim10^{13}$~\Mo, for a constant star formation rate over these first $\sim3$~Gyr.
In addition to the total stellar mass, a steeper IMF would affect
the stellar contribution to $Q_{\rm total}$, for instance, $\alpha=-2.65$ for $1 < \Mom\ < 10$ stars \citep{bcm10},
$\alpha=-5$ for $25 < \Mom\ < 120$ stars in the Magellanic Clouds \citep{mas02} or $\alpha=-2.45 \text{ to } -2.85$ in
external galaxies \citep{umm07,bml+15,wjf+15}. We therefore show the total ionising photon for a range of total
masses and IMF indices in Table~\ref{masses}.
\begin{table*}
  \caption{The total ionising photon rates, $Q_{\rm total}$, for various stellar
mass distributions. $M_{\rm max}$ is the maximum stellar mass permitted by limiting the total star formation rate to  SFR$_{\rm total}=1\times10^4$~\Moy\ (no AGN contribution to the FIR emission) and  $N_{\rm stars}$ is the number of stars of this mass.}
\centering
\begin{tabular}{l | ccc ccc ccc} 
\hline\hline
\smallskip
$M_{\rm total}$ [\Mo] & \multicolumn{3}{c}{$4\times10^{11}$} &  \multicolumn{3}{c}{$4\times10^{12}$} & \multicolumn{3}{c}{$4\times 10^{13}$}\\
 IMF index              & $\alpha=-2.35$  & $\alpha=-2.60$& $\alpha=-2.85$& $\alpha=-2.35$ & $\alpha=-2.60$& $\alpha=-2.85$& $\alpha=-2.35$ & $\alpha=-2.60$& $\alpha=-2.85$\\
\hline\hline
$M_{\rm max}$  [\Mo]  & 43                          & 110                   & 550                         & 6                    &9          & 17                                  & 1   & 1& 1\\
$N_{\rm stars}$      &$5\times10^{6}$     &$1\times10^{5}$ & 400 &  $5\times10^{9}$  &  $1\times10^{9}$  &$7\times10^{7}$&$4\times10^{12}$ &$3\times10^{12}$ & $2\times10^{12}$\\
$Q_{\rm total}$  [s$^{-1}$] & $9\times10^{56}$ &  $2\times10^{57}$ & $4\times10^{62}$ &  $2\times10^{55}$ &  $8\times10^{55}$ &  $3\times10^{56}$ &  $1\times10^{47}$ & $2\times10^{47}$ & $4\times10^{47}$\\
\hline
\end{tabular}
\label{masses}  
\end{table*}

Since the total stellar mass is dominated by the more numerous, least massive, cooler stars, ionising photon rates above
the critical 21-cm value favour $M_{\rm total} \lapp4\times10^{11}$ \Mo, although it
should be borne in mind that SFR$_{\rm total}=1\times10^4$~\Moy\ is most likely an upper limit, due to an AGN
contribution to the FIR luminosity \citep{msk+10,nrw+10}. For example, for a 50\% contribution,
the critical ionising photon rate is reached by just two of the canonical Galactic models ($M_{\rm
  total}=4\times10^{11}$~\Mo, Table~\ref{masses2}).
\begin{table*}
  \caption{As Table \ref{masses}, but limiting the total star formation rate to  SFR$_{\rm total}=5\,000$~\Moy\ (50\% AGN contribution to the FIR emission).}
\centering
\begin{tabular}{l | ccc ccc ccc} 
\hline\hline
\smallskip
$M_{\rm total}$ [\Mo] & \multicolumn{3}{c}{$4\times10^{11}$} &  \multicolumn{3}{c}{$4\times10^{12}$} & \multicolumn{3}{c}{$4\times 10^{13}$}\\
 IMF index              & $\alpha=-2.35$  & $\alpha=-2.60$& $\alpha=-2.85$& $\alpha=-2.35$ & $\alpha=-2.60$& $\alpha=-2.85$& $\alpha=-2.35$ & $\alpha=-2.60$& $\alpha=-2.85$\\
\hline
$M_{\rm max}$  [\Mo]  & 23                          &51                   & 190                         & 3                    &4          & 6                                 & $0.7$   & $0.6$& $0.6$\\
$N_{\rm stars}$      &$2\times10^{7}$     &$1\times10^{6}$ & 8000 &  $4\times10^{12}$  &  $7\times10^{9}$  &$1\times10^{9}$&$1\times10^{13}$ &$9\times10^{12}$ & $8\times10^{12}$\\
$Q_{\rm total}$  [s$^{-1}$] & $3\times10^{56}$ &  $4\times10^{56}$ & $3\times10^{53}$ &  $1\times10^{54}$ &  $1\times10^{55}$ &  $3\times10^{56}$ &  $1\times10^{43}$ & $2\times10^{43}$ & $1\times10^{43}$\\
\hline
\end{tabular}
\label{masses2}  
\end{table*}
This still begs the question of how the continual star formation is fuelled, although the stellar contribution to $Q_{\rm total}$ will
decrease further with an increasing AGN contribution. Furthermore, we have not accounted for shielding by dust nor how much
this attenuates the observed UV flux.  Also, while the critical $Q_{\rm total} = 3\times10^{56}$~s$^{-1}$ is
sufficient to ionise all of the neutral atomic gas in the Milky Way (where $n\lapp10$~\ccm, \citealt{kk09}), since
$Q\propto n^2$ \citep{ost89}, denser gas (e.g. in molecular clouds where $n\gapp10^{3}$~\ccm) requires much higher
ionising photon rates \citep{rjbg15}. Simulations of a stellar population distributed within a galactic disk would be
required in order to determine whether a distribution of UV luminous point sources would result in a Str\"{o}mgren
sphere of ``infinite'' radius, as is the case for a single centrally located $Q \gapp10^{56}$~s$^{-1}$ ionising source
\citep{cw12}.

\begin{figure}
\centering \includegraphics[angle=-90,scale=0.52]{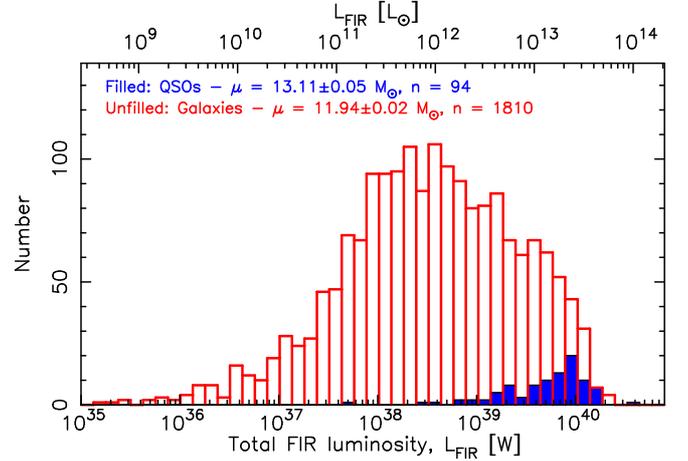}
\caption{The FIR luminosity distribution of the known QSOs and galaxies of the sample.}
\label{FIR-histo}
\end{figure}

Lastly, as Fig.~\ref{temp-bolo} shows, many of the highest FIR luminous sources are classified as QSOs, and so
known to host a powerful AGN. Although, forming only a small fraction (5\%) of the sources for which the FIR luminosity
can be derived (Fig.~\ref{FIR-histo}),
we see that the mean luminosity for the QSOs is an order of magnitude higher than for the galaxies
(as is also the case for the UV, Fig.~\ref{temp-Q}), lending support to the argument of an AGN contribution to
the FIR emission in the most luminous sources. There does remain, however, a  number of galaxies with these
same luminosities.


\section{Conclusions}

For the past decade evidence has been building of a critical photo-ionisation rate in the ultra-violet band, above which
all of the gas in the host galaxy is ionised. All of the observational evidence comes from searches of \HI\ 21-cm
absorption in redshifted radio sources, which has never been detected above rates of $Q=2.5\times10^{56}$ s$^{-1}$
($L_{\rm UV}\gapp10^{23}$~\WpHz, \citealt{cww+08,chj+19}). This observational result is supported by a model of a quasar
placed within an exponential gas disk, for which $Q\gapp3\times10^{56}$~s$^{-1}$ is sufficient to ionise all of the
neutral atomic gas in a large spiral \citep{cw12}. However, current 21-cm absorption searches are insensitive to column
densities of $N_{\rm HI}\ll 10^{20}$ atoms~\scm, and so the observations cannot rule out that the gas is merely heated
or partially ionised to below the detection limit.  Furthermore, the quasars with $L_{\rm UV}\gapp10^{23}$~\WpHz, tend
to be type-1 objects, implying a direct view to the naked AGN, whereas at $L_{\rm UV}\lapp10^{23}$~\WpHz, both type-1
(unobscured) and type-2 (obscured AGN) objects exhibit a 50\% detection rate for 21-cm absorption \citep{cw10}. This
would suggest that the non-detection of \HI\ above the critical UV luminosity is primarily an orientation effect,
although it would mean that the low luminosity type-1 AGN are somehow different from their high luminosity counterparts.

Thus, it is of great interest to confirm the possibility of a critical ionising luminosity in another band.
\citet{psv+12} report a critical X-ray luminosity ($L_{\mathrm{X}}\sim10^{37}$~W) above which 250~$\mu$m emission is not
detected in $1 < z < 3$ SPIRE sources, leading to the conclusion that the star formation is suppressed in these
objects. Since 21-cm absorption traces the reservoir for star formation, if the FIR emission is due mainly to stellar
heated dust, we may also expect a critical UV luminosity above which the 250~$\mu$m emission is suppressed.  By matching
the SPIRE sources to their NED counterparts, we obtain 14\,457 extragalactic objects with spectroscopic redshifts. Of
these, there is sufficient UV photometry to determine the ionising photon rate for 3315, and of the FIR detections
(i.e. detected above the SPIRE confusion limit of $3\sigma_{\rm conf}=18$ mJy) 2013 sources for which a modified
blackbody could be fit, yielding the dust temperature. From these, we find:

\begin{itemize}

\item A mean dust temperature of $\left<T_{\rm dust}\right> = 31.4\pm0.2$~K and a mean spectral emissivity index of
  $\left<\beta\right>=0.991\pm0.001$.

 \item No apparent critical ionising photon rate, with 250~$\mu$m emission being detected up to $Q=8.9\times10^{57}$
  s$^{-1}$. If the 21-cm results and model are reliable,  this suggests that the FIR emission does not exclusively trace the stellar heated dust,
implying a significant contribution from an AGN.

\item A strong correlation between the dust temperature and the ionising photon rate, which we suspect is driven mainly
  by the Malmquist bias. 
 Normalising the ionising photon rate by the total FIR luminosity, causes the
  correlation to disappear. 
  Since the number of  luminous AGN is expected to increase with redshift, this may also suggest  that the low temperature (FIR) emission  can arise from AGN heating of the dust.
\end{itemize}

By calculating the ionising photon rate expected for each stellar mass, for a total star formation rate of
SFR$_{\rm total}=1\times10^4$~\Moy, the observed ionising photon rates ($Q\sim10^{57}$ s$^{-1}$) in the most luminous of the
SPIRE sources can be reproduced by several initial mass functions, which give a sufficient population of massive stars.
Exceeding the critical value above which all of the neutral atomic gas is believed to be ionised, this raises
the question of what fuels the star formation. However, this is very much dictated by the choice of SFR$_{\rm total}$ and 
if $<10^4$~\Moy, due to a large AGN contribution to the FIR luminosity, lower values of $Q_{\rm total}$ are obtained. 
In order to address this, a valid IMF for such highly luminous sources is required to determine whether
the stellar population can account for the observed ionising photon rates. If the case, in the model
of \citet{cw12} the single large ionising source should be replaced with the putative stellar distribution 
embedded in the exponential gas disk, in order to determine whether such a population could completely ionise the gas.

\section*{Acknowledgements}
We wish to thank the anonymous referee for their prompt and helpful comments.
SWD acknowledges receipt of a Victoria Doctoral Scholarship and an Australian Government Research Training Program
  scholarship administered through Curtin University.  This research has made use of {\sc Astropy}, a
  community-developed core Python package for Astronomy (Astropy Collaboration, 2013), the NASA/IPAC Extragalactic
  Database (NED) which is operated by the Jet Propulsion Laboratory, California Institute of Technology, under contract
  with the National Aeronautics and Space Administration. This research has also made use of NASA's Astrophysics Data
  System Bibliographic Services.


\begin{appendix} 

\section{UV photo-ionisation of the Chandra Deep Field North sources}
\label{sec:x-ray}

The original aim of the project was to test the \citet{psv+12} sample for a similar
absence of 250~$\mu$m emission above the critical UV luminosity where \HI\ 21-cm absorption 
is not detected. As an initial step we reproduced the method of \citet{psv+12}, cross-matching
the {\em Chandra Deep Field North} (CDF-N, \citealt{abb+03}) with spectroscopic redshifts, obtained from
\citet{bcw08,tbc+08}, as well as the references listed in table S2 of \citeauthor{psv+12}.
For each source for which we could obtain a redshift, we calculated the \emph{k}-corrected X-ray
luminosity, $L_{\mathrm{X}}$, from the flux density, assuming $S_{\nu} \propto \nu^{-0.9}$. 
We then cross-matched these with the sources within 6 arc-sec of those searched by HerMES. 
 These matches are unique, i.e., no sources within SPIRE are matched to multiple CDF-N AGN. 

\begin{figure}
\centering \includegraphics[angle=-90,scale=0.50]{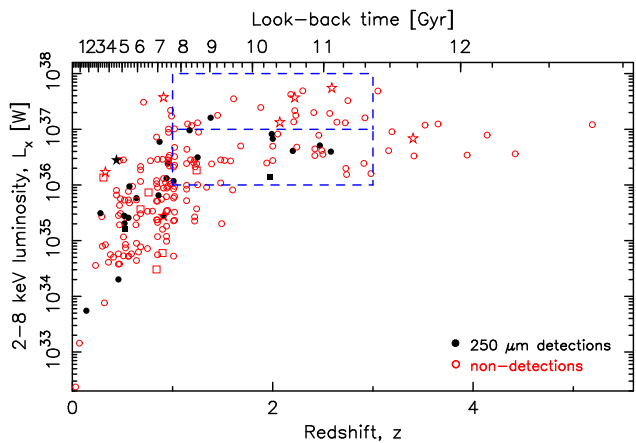}
\caption{The 2--8 keV X-ray luminosity versus redshift for the sources in the CDF-N searched in 250~$\mu$m by SPIRE.  The broken lines enclose the $L_{\mathrm{X}} = 10^{43} - 10^{44}$ erg s$^{-1}$ and $L_{\mathrm{X}} = 10^{44} - 10^{45}$ erg s$^{-1}$ regions
  spanning $z=1-3$ (cf. figure~1 of \citealt{psv+12}).  The filled symbols show those detected in 250~$\mu$m emission
  ($S_{\rm FIR} > 18$ mJy) and the unfilled those undetected, with the shapes representing the NED classification -- stars
  for QSOs, circles for galaxies and squares for unclassified.}
\label{psv_z}
\end{figure}

In Fig. \ref{psv_z}, we show the resulting 2--8 keV X-ray luminosity versus redshift for the 250~$\mu$m searched sources. 
This is identical to the distribution of \citet{psv+12}, above the $L_{\mathrm{X}} < 10^{42}$erg s$^{-1}$ ($<
10^{35}$~W) cut, apart from an additional source at $z=1.3709$ (source number 40 in \citealt{tbc+08}), which has
$L_{\mathrm{X}} = 1.6\times10^{37}$~W and $S_{\rm FIR}= 40.1$~mJy.\footnote{This source, SDSS\,J123553.13+621037.3, has
  a $0.64$ arc-sec offset from the optical offset and a $5.9$ arc-sec offset from the 250~$\mu$m source and so just
  falls within the 6 arc-sec search radius.}  Examining the statistics, below the $L_{\mathrm{X}} =10^{37}$~W cut-off, there
are 10 FIR detections and 34 non-detections, giving a detection probability of $p=0.227$. Applying this to the
$L_{\mathrm{X}} >10^{37}$~W sample, the binomial probability of obtaining zero 250~$\mu$m detections in 21 sources is
$P({\rm bin}) =4.485\times10^{-3}$. This is significant at $S({\rm bin})=2.84\sigma$ and
consistent with \citeauthor{psv+12} finding a $> 99\%$ ($>2.58\sigma$) significance from a single-tail Fisher's exact
test.  This significance does, however, decrease when source
number 40 is included or the whole sample tested (Table \ref{psv_stats}).
\begin{sidewaystable} 
\centering
 \caption{The 250~$\mu$m detection statistics above and below $L_{\mathrm{X}} = 10^{44}$ erg s$^{-1}$ ($10^{37}$~W) for
    the $L_{\mathrm{X}} =10^{37} - 10^{39}$ W---$z=1-3$ sample, the $L_{\mathrm{X}} >10^{35}$~W ($> 10^{42}$erg
    s$^{-1}$) sample and the whole sample (\citealt{psv+12}), in addition to the effect on each by including source
    number 40. $p$ gives the probability of a detection based upon the detection rate at $L_{\mathrm{X}} <10^{37}$ W,
    with the last two columns giving the binomial probability and significance of this observed number of 250~$\mu$m
    detections or fewer  at $L_{\mathrm{X}} > 10^{37}$ W.}
\centering
\begin{tabular}{l c  c c c  c c  c c c} 
\hline\hline
\smallskip
Sample &  \multicolumn{3}{c}{$L_{\mathrm{X}} < 10^{37}$ W} &  Source & $n_{\rm total}$ & \multicolumn{2}{c}{$L_{\mathrm{X}}\geq10^{37}$ W }& $P({\rm bin})$ & $S({\rm bin})$\\ 
             & Non-detections &  Detections &  $p$  &  No. 40 & & Non-detections &  Detections &  & \\
\hline
$L_{\mathrm{X}} = 10^{36} - 10^{38}$ W         &  34   & 10 & 0.227  & Yes  &66 &   21 & 1  & 0.0257 & $2.23\sigma$\\
                                  &         &       &          &  No   & 65 &  21 & 0  & $4.485\times10^{-3}$ & $2.84\sigma$\\
$L_{\mathrm{X}} > 10^{35}$ W  & 122  &  23  &  0.159 & Yes & 176 & 30  & 1 & 0.0325 & $2.14\sigma$\\
                                 &         &       &          &  No   & 175 & 30 & 0  & $5.620\times10^{-3}$ & $2.77\sigma$\\
Whole  sample   & 149 &  24 &  0.139 & Yes & 205 & 30  & 1 & 0.0585 & $1.89\sigma$\\
                            &         &       &                                    &  No   & 204 & 30 & 0  & 0.0113 & $2.53\sigma$\\
\hline
\end{tabular}
\label{psv_stats}  
\end{sidewaystable}  
Thus, the absence of 250~$\mu$m emission in luminous X-ray sources is not particularly significant, especially compared
to $S({\rm bin}) = 5.66\sigma$ for the absence of 21-cm absorption in the UV luminous sources
\citep{chj+19}.

\begin{figure}
\centering \includegraphics[angle=-90,scale=0.51]{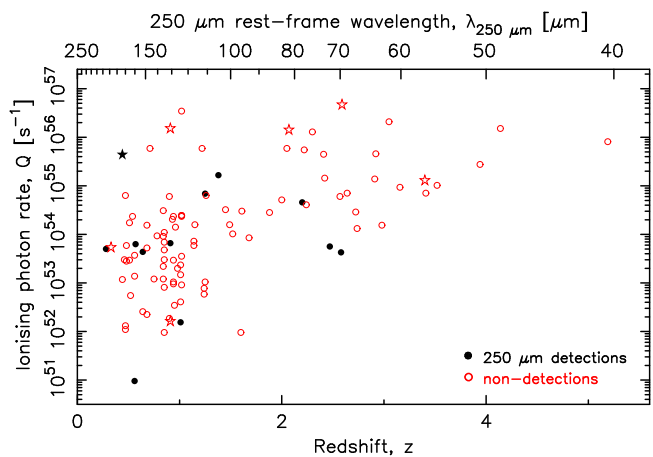}
\caption{The ionising photon rate versus redshift for the sources in the CDF-N. The symbols are as per
  Fig. \ref{psv_z}.} 
\label{Q_250}
\end{figure}

Following the SED fitting procedure described in Sect.~\ref{sec:pf}, there is sufficient photometry to obtain the
ionising photon rate for 103 of the CDF-N sources (Fig.~\ref{Q_250}).
The highest ionising photon rate at which 250~$\mu$m is detected is $Q=4.4\times10^{55}$~s$^{-1}$, which is
consistent with the value above which \HI\ 21-cm absorption remains undetected (Sect.~\ref{intro}).  However, the
$p=0.134$ detection rate below this gives a binomial probability of $P({\rm bin}) = 0.137$ for zero detections out of
the 14 sources with $Q>4.4\times10^{55}$~s$^{-1}$, which is only significant at $S({\rm bin}) = 1.51\sigma$.
Moreover, when tested over a larger sample no critical rate is apparent (Sect.~\ref{cipr}).  Furthermore, extending the \citeauthor{psv+12} sample, using the {\em Chandra Deep Field South}
(CDF-S, \citealt{xlb+11}) and COSMOS fields \citep{ecv+09}, finds no evidence for suppressed star formation at high
X-ray luminosities \citep{ham+12}.
\end{appendix}

\end{document}